\documentstyle[preprint,revtex]{aps}
\begin{document}
\preprint{UM-P-92/69}
\preprint{OZ-92/21}
\preprint{July 1992}
\begin{title}
Topological phase due to electric dipole moment\\
and magnetic monopole
interaction
\end{title}
\author{Xiao-Gang He and Bruce H.J. McKellar}
\begin{instit}
Research Center for High Energy Physics\\
School of Physics\\
University of Melbourne \\
Parkville, Vic. 3052 Australia
\end{instit}
\begin{abstract}
We show that there is an analogous Aharonov-Casher topological effect on
a neutral particle with electric dipole moment interacting with a magnetic
field produced by magnetic monopoles.
\end{abstract}
\newpage
In 1984 Aharonov and Casher(AC) \cite{1} showed that there is a topological
effect on
a neutral particle with magnetic dipole moment when the particle is moving
on a plane under the influence of an electric field pruduced by a uniformly
charged infinitly long filament perpendicular to the plane. When the particle
moves around the filament, the wave function will develop a phase
equal to $\mu \lambda_e$. Here $\mu$ is the magnetic dipole moment of the
particle and $\lambda_e$ is the linear electric charge density of the filament.
In the configuration discussed here, the particle is classically force free.
This effect is called the AC effect. Several groups have
considered this effect\cite{2,3}. A more general derivation and the specific
conditions for the AC effect is given in Ref.\cite{2}. The AC effect has also
been
observed experimentally using neutrons\cite{4}.

In this comment we remark that should magnetic monopoles exist, the dual of AC
effect would be possible, namely, a neutral particle with
an electric dipole moment $d$ will experience a topological effect due to the
magnetic field ${\bf B}$ produced by magnetic monopoles. The
equation of motion for a spin half neutral particle with an electric dipole
moment is
\begin{equation}
(\not\partial + m + i{d\over 2}\sigma_{\mu\nu}\gamma_5F^{\mu\nu})\psi = 0\;.
\end{equation}
Assuming that there is only a static magnetic field ${\bf B}$, this becomes
\begin{equation}
(\not\partial + m + id {\bf \sigma}\cdot {\bf B}\gamma_5)\psi = 0\;,
\end{equation}
which can be further rewritten as
\begin{equation}
(\not\partial + m - d{\bf \gamma}\cdot {\bf B}\gamma_4)\psi = 0\;.
\end{equation}

This equation has the same form as the equation of motion for a neutral
particle with a magnetic dipole moment $\mu$ moving in a static electric
field ${\bf E}$. Indeed changing $-d$ to $\mu$ and ${\bf B}$ to ${\bf E}$ we
obtain the equation of motion for the later. We imediatly anticipate
that equation (2) implies the existence of a topological phase for the particle
wavefunction when appropriate conditions
are satisfied. One of the conditions for topological AC phase in the wave
function
is that there are regions where ${\bf \bigtriangledown }\cdot {\bf E}\not=0$.
In our case one would require that there are regions where ${\bf
\bigtriangledown}\cdot {\bf B}$ is not zero. But in the Maxwellian
electrodynamics the magnetic field ${\bf B}$
is always divergence-less, and one would not have any topological phase.
However, if there are Dirac magnetic monopoles one will
have ${\bf \bigtriangledown}\cdot {\bf B} = \rho_m$, where $\rho_m$ is magnetic
monopole density.  Although magnetic monopoles have not been observed it is
interesting to speculate along these lines.
In the following we will
determine the conditions for developing topological phase in
the wave function.

Our discussion is modeled on that for the AC effect in Ref.\cite{2}. The basic
idea is to find a solution of
equation (1) which can be writen as
\begin{equation}
\psi = exp[-id{\bf a}\gamma_4\int^{\bf r} {\bf \Gamma}\cdot d{\bf l}] \psi'\;,
\end{equation}
where $\psi'$ is a solution of free Dirac equation of motion, the matrix ${\bf
a}$ and
the field ${\bf \Gamma}$ are determined by matching equation(1) with
equation(4). One also must
make sure that ${\bf \bigtriangledown}\times {\bf \Gamma} =0$ in the region of
space where equation (4) holds. We find
\begin{equation}
-d{\bf \gamma}\cdot {\bf B} = i{\bf \gamma}\cdot {\bf \Gamma}{\bf a};\;\;
{\bf a}\gamma_4\gamma_\mu = \gamma_\mu{\bf a}\gamma_4.
\end{equation}
These equations have consistent solutions only
in two or less dimensions (spatial). The one dimensional solution
is of no interesting to us. So we must consider the motion of the particle move
in
a plane. Let the plane be the x-y plane. We find ${\bf a} = \sigma_{12}$,
${\bf \Gamma} = (-B_y, B_x, 0)$ and on this plan $\partial _z B_z = 0$.
In the ${\bf \bigtriangledown}\cdot {\bf B}=0$ region, we have
\begin{equation}
\psi = exp[id\sigma_{12}\gamma_4\int^{\bf r}({\bf B}\times {\bf z})\cdot d{\bf
l}]\psi'\;,
\end{equation}
where ${\bf z}$ is the unit vector along z-direction.
In this region ${\bf \bigtriangledown}\times {\bf \Gamma}$ does vanish.

If we let the particle to move a closed path in a plane in the
${\bf \bigtriangledown}\cdot {\bf B}=0$ region which encloses some region where
${\bf \bigtriangledown}\cdot {\bf B}$ is not zero, the wave function of the
particle
will develop a phase
\begin{equation}
\phi = -d\lambda_m\;,
\end{equation}
where $\lambda_m$ is the linear magnetic monopole density.
The phase is respect to $\psi'$. We have shown that there is an analogous AC
effect as stated earlier.

There is a 1.5 standard deviation discrepency between the experimental value
and the theoretical
prediction for the AC effect. $\phi_{exp}/\phi_{theo} = 1.46 \pm 0.35$ and
$\phi_{exp} = 2.19 \pm 0.52 \; mrad$. While we would not wish to encourage the
idea that this discrepency is more than an indication that this difficult
experiment should be repeated, we have checked on its implications for the
present
effect. Taking into account the limit on the electric dipole moment of the
neutron ($d < 1.2 \times 10^{-25} ecm$ \cite{5}), the linear magnetic monopole
density would have to be at least $10^{22}/ecm$ to bring theory and experiment
into agreement. Needless to say, such large magnetic monopole densities are
ruled out by other experiments\cite{6}.

\acknowledgments
XGH would like to thank A. Davies, A. Klein and G. Opat for useful discusions.
This work is supported in part by the Australian Research Council. We thank the
Department of Energy's Institute for Nuclear Theory at the University of
Washington for hospitality and the Department of Energy for partial support
during the completion of this work.

\end{document}